\documentclass[10pt,aps,prb,twocolumn]{revtex4-2}
\usepackage[english]{babel}
\usepackage[T1]{fontenc} 
\usepackage[utf8]{inputenc} 
\usepackage{verbatim}
\usepackage{graphicx}
\usepackage{graphics}
\usepackage{color}
\usepackage{textcomp}
\usepackage{amsfonts}
\usepackage{amsmath}
\usepackage{hyperref}
\usepackage{amssymb}
\usepackage{mathtools}
\usepackage{tabularx}
\usepackage[a4paper, margin=1.5cm]{geometry}

\newcolumntype{Y}{>{\centering\arraybackslash}X}

\begin{document}

\title{Supervised and Unsupervised Deep Learning Applied to the Majority Vote Model}

\author{J. F. Silva Neto}
\affiliation{Departamento de Matem\'{a}tica e F\'{\i}sica - Campus Caxias, Universidade Estadual do Maranh\~{a}o, 65604-380, Caxias - MA, Brazil}
\author{D. S. M. Alencar}
\affiliation{Departamento de Matem\'{a}tica e F\'{\i}sica - Campus Caxias, Universidade Estadual do Maranh\~{a}o, 65604-380, Caxias - MA, Brazil}
\author{L. T. Brito}
\affiliation{Departamento de F\'{i}sica, Universidade Estadual do Piau\'{i}, 64002-150, Teresina - PI, Brazil}
\author{G. A. Alves}
\affiliation{Departamento de F\'{i}sica, Universidade Estadual do Piau\'{i}, 64002-150, Teresina - PI, Brazil}
\author{F. W. S. Lima}
\affiliation{Departamento de F\'{\i}sica, Universidade Federal do Piau\'{i}, 57072-970, Teresina - PI, Brazil}
\author{A. Macedo-Filho}
\affiliation{Departamento de F\'{i}sica, Universidade Estadual do Piau\'{i}, 64002-150, Teresina - PI, Brazil}
\author{R. S. Ferreira}
\affiliation{Departamento de Ci\^{e}ncias Exatas e Aplicadas, Universidade Federal de Ouro Preto, 35931-008, Jo\~{a}o Monlevade - MG, Brazil}
\author{T. F. A. Alves}
\affiliation{Departamento de F\'{\i}sica, Universidade Federal do Piau\'{i}, 57072-970, Teresina - PI, Brazil}

\date{Received: date / Revised version: date}

\begin{abstract}

We employ deep learning techniques to investigate the critical properties of the continuous phase transition in the majority vote model. In addition to deep learning, principal component analysis is utilized to analyze the transition. For supervised learning, dense neural networks are trained on spin configuration data generated via the kinetic Monte Carlo method. Using independently simulated configuration data, the neural network accurately identifies the critical point on both square and triangular lattices. Classical unsupervised learning with principal component analysis reproduces the magnetization and enables estimation of critical exponents, typically obtained via Monte Carlo importance sampling. Furthermore, deep unsupervised learning is performed using variational autoencoders, which reconstruct input spin configurations and generate artificial outputs. The autoencoders detect the phase transition through the loss function, quantifying the preservation of essential data features. We define a correlation function between the real and reconstructed data, and find that this correlation function is universal at the critical point. Variational autoencoders also serve as generative models, producing artificial spin configurations.

\end{abstract}



\maketitle

\section{Introduction}

Machine learning comprises a diverse set of techniques designed to analyze large-scale datasets (big data), enabling inference, classification, and prediction. The analysis of big data has become essential for various scientific disciplines, notably Statistical Physics~\cite{Mehta-2019, Rodrigues-2023}. Other major fields in Physics that benefit from big data analysis include Condensed Matter Physics, High Energy Physics, Astrophysics, and Cosmology. Additionally, quantum computing is an emerging area that increasingly incorporates machine learning approaches for data analysis~\cite{Mehta-2019, Rodrigues-2023}.

Machine learning is a subfield of artificial intelligence~\cite{Russell-2010, Chollet-2021} focused on developing algorithms that efficiently recognize patterns in datasets. Complex systems are thus treated as large collections of data. These algorithms learn from data, enabling models to make predictions, draw inferences, or adaptively perform tasks without explicit human intervention. The primary objective of machine learning algorithms is to uncover patterns within datasets. 

Machine learning algorithms are commonly classified into three principal categories~\cite{Murphy-2012}: supervised learning, unsupervised learning, and reinforcement learning. Each category possesses distinct features and serves specific purposes. Supervised learning utilizes labeled datasets, where each data instance is associated with a known output, allowing the model to learn explicit input-output mappings for prediction. Typical supervised learning tasks include regression and classification.

In contrast, unsupervised learning operates on unlabeled data, requiring the model to autonomously discover patterns or structures within the dataset. Common unsupervised algorithms include clustering methods such as \textit{K-means}, \textit{DBSCAN}, and hierarchical clustering (\textit{HCA}); anomaly detection techniques like support vector machines (\textit{SVM}) and isolation forests; and dimensionality reduction approaches such as principal component analysis (\textit{PCA}) and t-distributed stochastic neighbor embedding (\textit{t-SNE}).

The third major group, reinforcement learning algorithms, differs fundamentally from the previous two. In reinforcement learning, the model acts as an \textbf{agent} that interacts with its environment by selecting actions and receiving feedback in the form of rewards or penalties. The objective is that the agent autonomously learn an optimal strategy, known as the \textbf{policy}, that maximizes rewards while minimizing penalties.

Our primary objective in this work is to employ supervised and unsupervised learning algorithms~\cite{Murphy-2012} to investigate the continuous phase transitions in the majority vote (MV) model~\cite{Oliveira-1992, Pereira-2005, Wu-2010, Yu-2017, Crochik-2005, Vilela-2009, Lima-2012, Vieira-2016, Fronczak-2017, Lima-2006, Lima-2007, Vilela-2018, Krawiecki-2018, Krawiecki-2018-2, Alves-2019, Krawiecki-2019, Vilela-2020-2, Zubillaga-2022, Alencar-2023, Alencar-2024}. The MV model is an irreversible, nonequilibrium kinetic spin system that exhibits a continuous phase transition belonging to the Ising universality class.

Defined on a regular lattice, each site $i$ hosts a spin variable $\sigma_i = \pm 1$. This model serves as a paradigm for consensus formation, where individuals (spins) tend to align with the majority of their neighbors, subject to a noise parameter $q$ that introduces a probability of dissent. At low noise levels, the system favors consensus, whereas at high noise levels, it remains disordered. The critical point $q_c$ separates the ordered and disordered phases.

Some works have applied machine learning techniques to study equilibrium and nonequilibrium phase transitions. For instance, Carrasquilla and Melko~\cite{Carrasquilla-2017} employed supervised learning with neural networks to classify phases in the two-dimensional Ising model. They trained a dense neural network on spin configurations generated via Monte Carlo simulations at various temperatures, enabling the network to learn the distinction between ordered and disordered phases. The trained network successfully identified the critical temperature and estimated critical exponents.

We also highlight related approaches applied to directed percolation~\cite{Shen-2021}, the pair contact process with diffusion~\cite{Shen-2022}, and various extensions of the Ising model~\cite{Tola-2023}. Classification neural networks have proven effective for determining critical thresholds, including in the context of quantum phase transitions~\cite{vanNieuwenburg-2017}. A comprehensive investigation into the minimal neural network architecture necessary for phase classification is presented in Ref.~\cite{Kim-2018}. Unsupervised learning methods have likewise been utilized to explore phase transitions in the Ising model~\cite{Wetzel-2017, Mehta-2019, Walker-2020}. In this work, we extend both supervised and unsupervised learning techniques to the MV model on square and triangular lattices.

Although our focus is on supervised and unsupervised learning techniques applied to the critical behavior of magnetic systems, it is important to highlight the technological and strategic significance of reinforcement learning. This approach underpins the development of advanced personal assistant models such as \textit{ChatGPT} and has driven major advances in automation and robotics~\cite{Geron-2019}. Additionally, reinforcement learning is increasingly relevant in quantum computing applications~\cite{Rodrigues-2023}. In the following, we present a concise overview of machine learning techniques, emphasizing deep learning methods.

Machine learning problems generally share a common structure. We begin with an observable $\boldsymbol{y}$, which depends on independent variables $\boldsymbol{x}$. The dataset is denoted by $\mathcal{D} = (\boldsymbol{X}, \boldsymbol{y})$, comprising the observable data $\boldsymbol{y}$ as a function of the matrix of independent variables $\boldsymbol{X}$. The next step is to select a model $f\left(\boldsymbol{X}; \boldsymbol{\theta} \right)$ (also referred to as a predictor), typically a function $f:\boldsymbol{x} \to \boldsymbol{y}$ parameterized by a set of hyperparameters $\boldsymbol{\theta}$. The model enables predictions of new instances $\boldsymbol{y}'$ of the observable $\boldsymbol{y}$
\begin{equation}
   \boldsymbol{y}' = f\left(\boldsymbol{X}; \boldsymbol{\theta} \right).
   \label{predictor}
\end{equation}

The training process of the model aims to determine the optimal set of hyperparameters $\boldsymbol{\hat{\theta}}$ that best fit the data. Typically, this is achieved by minimizing a cost or loss function:
\begin{equation}
   E(\boldsymbol{y},\boldsymbol{y}') = E\left[\boldsymbol{y}, f\left(\boldsymbol{X}; \boldsymbol{\theta} \right) \right],
   \label{cost-function}
\end{equation}
which quantifies the discrepancy between the actual data $\boldsymbol{y}$ and the model predictions $\boldsymbol{y}'$. The optimal parameters $\boldsymbol{\hat{\theta}}$ are found by solving the minimization problem
\begin{equation}
   \boldsymbol{\hat{\theta}} = \underset{\boldsymbol{\theta}}{\mathrm{argmin}}\, E\left[\boldsymbol{y}, f\left(\boldsymbol{X}; \boldsymbol{\theta} \right) \right].
   \label{minimization}
\end{equation}

Each application of machine learning encounters common challenges, foremost among them the quality and quantity of available data. When trained on large and representative datasets, different models tend to exhibit comparable behavior and predictive accuracy~\cite{Banko-2001}. Limitations in data quantity can often be mitigated through augmentation techniques. For instance, in spin systems on a square lattice, one can generate additional configurations by rotating the lattice by $90^{\circ}$ or by inverting all spins to obtain new valid spin configurations.

A fundamental example of a machine learning algorithm is the least squares method, which fits a linear function $\boldsymbol{y} = a\boldsymbol{x} + b$ to sparse data~\cite{Geron-2019} by minimizing the mean squared error (MSE)
\begin{equation}
  \ell_{\text{MSE}} = \frac{1}{N_{\mathcal{D}}} \sum^{N_{\mathcal{D}}}_{i=1} \left[y_i - y^\prime_i(\boldsymbol{\theta})\right]^2,
  \label{loss-mse}
\end{equation}
where $N_\mathcal{D}$ denotes the total size of the dataset $\mathcal{D}$. In deep learning, the neural network acts as the model $f\left(\boldsymbol{X}; \boldsymbol{\theta} \right)$, but these are nonlinear models characterized by numerous hyperparameters. Neural networks are inspired by the structure of biological neurons in the nervous system. In a playful analogy, neural networks may be regarded as regression methods ``on steroids''.

The fundamental unit of neural networks is the neuron. A neuron receives as input a vector of $d$ features, $\boldsymbol{x} = (x_1, x_2, \ldots, x_d)$, and produces a scalar output $h_i(\boldsymbol{x})$. The output is generated in two steps. First, a linear transformation maps the inputs to a scalar $g_i$:
\begin{equation}
   g_i = \boldsymbol{\omega}_i \cdot \boldsymbol{x} + b_i,
   \label{neuron-linear-1}
\end{equation}
where $\boldsymbol{\omega}_i$ are the weights associated with neuron $i$, and $b_i$ is its bias. The second step introduces nonlinearity by using an activation function $\varrho(x)$, yielding the final neuron output
\begin{equation}
   h_i = \varrho(g_i).
   \label{neuron-output}
\end{equation}
If $\varrho(x)$ is a linear function, such as $\varrho(x) = ax$, the neuron reduces to a purely linear transformation of the input features. 

A neural network is then composed of stacked layers of neurons, where the output of each layer serves as the input to the subsequent layer. The layer that receives the data features is called the input layer, the intermediate layers are called hidden layers, and the final layer is the output layer. When a neural network contains multiple layers, it is referred to as a deep neural network. 

In this way, a neural network is a nonlinear model with numerous hyperparameters, which are the weights and biases of each of its neurons. Thanks to the large number of hyperparameters and the inherent nonlinearity, neural networks are capable of representing highly complex functions. Increasing the depth of the neural network enables it to capture more intricate features of the dataset.

It is important to note that neural networks are susceptible to the same challenges as regression models, such as underfitting and overfitting. The capacity of a neural network can be increased by adding more neurons per layer or by stacking additional layers, thereby making the network deeper. Selecting the optimal architecture for a given problem typically involves a process of trial and error, guided by intuition and experience. 

However, a central challenge remains: how to effectively train neural networks. The problem of training neural networks was resolved with the development of the backpropagation algorithm~\cite{Anderson-1988} which is essentially an adaptation of optimization techniques such as the gradient descent method (also known as the \textit{steepest descent} method~\cite{Arfken-2013}). Examples of state-of-the-art optimization algorithms include the \textit{Stochastic Gradient Descent} (\textit{SGD})~\cite{Robbins-1951}, Nesterov gradients~\cite{Nesterov-1983}, \textit{RMSprop}~\cite{Tieleman-2012}, and \textit{ADAM} (Adaptive Moment Estimation)~\cite{Kingma-2014}, etc.

The optimization method introduces a new parameter $\eta$, called the learning rate. The learning rate controls the step size taken during each iteration of the optimization process. A small learning rate may lead to slow convergence, while a large learning rate can potentially lead to divergence. Therefore, selecting an appropriate learning rate is crucial for effective training. 

Several obstacles can hinder successful training, depending on the specific nature of the problem. The most common issues are:
\begin{itemize}
   \item \textbf{Vanishing gradients}: Gradients in deeper layers approach zero, especially with sigmoid activation functions, preventing the backpropagation algorithm from effectively updating the parameters of these layers;
   \item \textbf{Exploding gradients}: Gradients become excessively large, impeding convergence of the model parameters;
   \item \textbf{Overfitting}: The model fits the training data too closely, often due to insufficient data. Neural networks, with their many hyperparameters, are particularly prone to overfitting and thus require large training datasets;
   \item \textbf{Slow convergence}: Training proceeds very slowly, which can be mitigated by tuning the optimization method or its parameters.
\end{itemize}

Many of these problems have established countermeasures. For vanishing gradients, which are common in hidden layers with sigmoid or hyperbolic tangent activations, switching to \textit{ReLU} or its variants (\textit{LeakyReLU}, \textit{ELU}, \textit{SELU}) can help. Another remedy for both vanishing and exploding gradients is the introduction of batch normalization~\cite{Ioffe-2015} during training.

Batch normalization centers and normalizes the inputs of each layer, introducing additional hyperparameters but accelerating convergence. It can also mitigate vanishing gradients in layers with sigmoid or hyperbolic tangent activations and acts as a regularizer for reasons not yet fully understood~\cite{Geron-2019}. The batch normalization effectively inserts an extra layer with weights $\boldsymbol{\omega^\prime_i}$ and biases $b^\prime_i$, which are updated via backpropagation.

To address overfitting, regularization algorithms such as $\ell_1$ and $\ell_2$ regularization can be applied to the outputs of the layers. These add a regularization term to the total cost function, each with distinct effects:
\begin{itemize}
   \item $\ell_1$ (or LASSO) regularization: Adds the norm of the neuron weights, scaled by a factor $\lambda$, to the error. This encourages sparsity, causing many weights to become zero during training (feature selection effect);
   \item $\ell_2$ (or RIDGE) regularization: Adds the squared norm of the neuron weights, also scaled by $\lambda$, to the error. This does not induce sparsity but encourages weights to cluster near zero.
\end{itemize}
Another important regularization technique is \textit{Dropout}~\cite{Srivastava-2014, Geron-2019}, which randomly deactivates a fraction of neurons during training to eliminate spurious correlations and accelerate convergence. The dropout rate controls the proportion of neurons deactivated in each training iteration.

In the following sections, we apply supervised and unsupervised learning techniques to study the continuous phase transition in the MV model on both square and triangular lattices. We begin by describing the MV model and the Monte Carlo method used to generate spin configurations. We then present our results using supervised learning with dense neural networks, followed by unsupervised learning with \textit{PCA} and also dense neural networks.

\section{Data Generation}

We describe the generation of spin configuration data for the MV model. First, we briefly introduce the model and the kinetic Monte Carlo simulation technique~\cite{Landau-2015}, which is used to obtain stationary spin configurations. After detailing the data collection methodology, we outline the machine learning techniques and training protocols employed to achieve our results.

\begin{figure}[!ht]
   \begin{center}
   \includegraphics[scale=0.3]{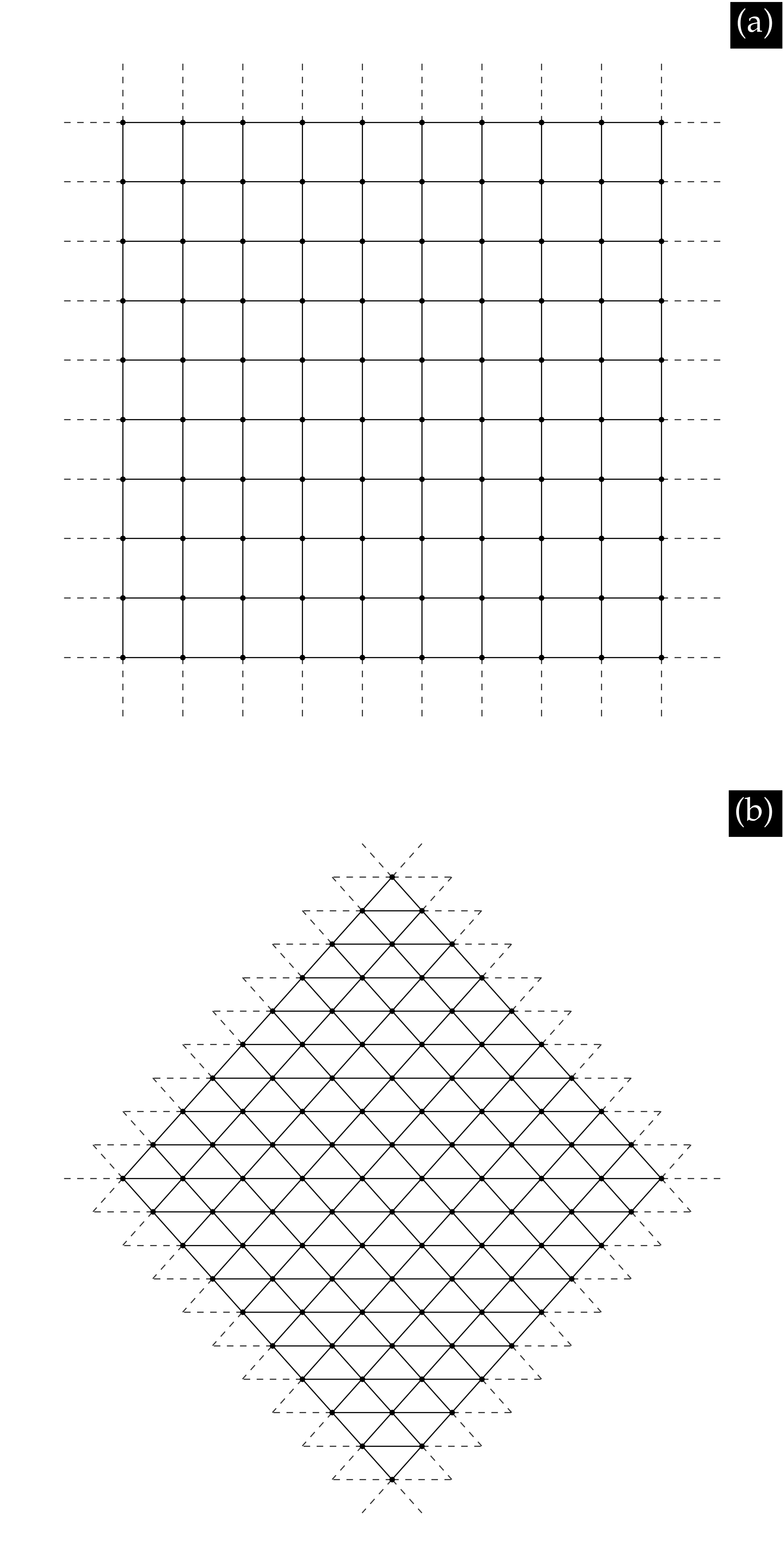}
   \end{center}
   \caption{Panel (a) illustrates a square lattice of size $L=10$ with periodic boundary conditions. Panel (b) displays a triangular lattice with identical parameters. Lattice nodes are depicted as dots, while edges represent nearest-neighbor interactions. Dashed edges indicate periodic boundary conditions, connecting nodes on opposite sides of the lattice.}
   \label{lattices}
\end{figure}

We consider the MV model on a square lattice, shown in panel (a) of Fig.~(\ref{lattices}) with $N=L^2$ nodes and periodic boundary conditions. The kinetic evolution rules of the MV model~\cite{Oliveira-1992, Pereira-2005, Yu-2017} are
\begin{enumerate}
   \item Each of the $N$ lattice nodes is assigned a spin variable $s_{i} = \pm 1$, representing two possible opinions (in favor or against). At each node, we associate a stochastic variable $s = \pm 1$, so the lattice state is
   \begin{equation}
      \boldsymbol{s} = \left( s_1, s_2, ..., s_N \right).
      \label{state}
   \end{equation}
   The dynamics starts with randomly selected opinion states for each node;
   \item At each step, a node $j$ is randomly chosen. The spin state is changed with probability $w_i(\boldsymbol{s})$, given by
      \begin{equation}
          w_i(\boldsymbol{s}) = \frac{1}{2} \left[1-(1-2q)s_{i}S\left(\sum_{\delta}^{z_{i}} s_{\delta} \right)\right],
      \label{mvfliprate}
      \end{equation}   
   where $q$ is the noise parameter, the summation index $\delta$ runs over the $z_i$ neighbors of spin $s_i$, and $S(x)$ is the sign function, which gives the majority opinion among the nearest neighbors:
   \begin{equation}
      S(x) = \begin{cases}
                   -1, & \text{if } x < 0; \\
                    0, & \text{if } x = 0; \\
                    1, & \text{if } x > 0.
               \end{cases}
   \end{equation}
\end{enumerate}
A change of opinion according to $w_i$ is equivalent to spin $s_i$ adopting the majority opinion of its neighbors with probability $1-q$, and the minority opinion with probability $q$. If there is no local majority, $s_i$ assumes either opinion with equal probability $w_i(\boldsymbol{s}) = 1/2$. The noise $q$ induces a continuous phase transition, analogous to the ferro-paramagnetic transition.

The following master equation describes the dynamics of the MV model~\cite{Landau-2015}:
\begin{equation}
   \frac{d}{dt} \mathcal{P}_{\boldsymbol{s}} = \sum_i^N w_i(\boldsymbol{s^i}) \mathcal{P}_{\boldsymbol{s^i}} - w_i(\boldsymbol{s}) \mathcal{P}_{\boldsymbol{s}},
   \label{masterequation}
\end{equation}
where $\mathcal{P}_{\boldsymbol{s}}$ is the statistical distribution of system states and
\begin{equation}
   \boldsymbol{s^i} = \left( s_1, s_2, ...,-s_i, ..., s_{N} \right),
   \label{flippedstate}
\end{equation}
is the state after one spin has been updated. The stationary solution of $\mathcal{P}_{\boldsymbol{s}}$ requires the local energies to obey the Boltzmann distribution.

One Monte Carlo step is defined as the sequential update of $L^2$ spins, which sets the simulation time unit. The system is allowed to reach a stationary state from a random initial configuration over $N_\text{term}$ thermalization steps. Once stationary, a time sequence of $N_t$ configurations $\boldsymbol{s_\ell}$ ($\ell = 0,1,...,N_t$) is collected. To avoid statistical correlations, additional Monte Carlo steps are discarded between subsequent stationary configurations~\cite{Landau-2015}.

To ensure that the system reached the stationary state, the initial $N_\text{term}$ Monte Carlo steps are discarded. After the system reaches the stationary regime, we record a time series consisting of $N_t$ configurations $\boldsymbol{s_\ell}$ ($\ell = 0,1,\ldots,N_t$). In order to suppress statistical correlations between successive configurations, additional Monte Carlo steps are discarded between consecutive elements of the recorded time series~\cite{Landau-2015}.

\section{Supervised Learning}

Using the Monte Carlo method, we generate the dataset used to train the model for classifying configurations as either ferromagnetic or paramagnetic. Supervised training requires some prior knowledge of the phase transition, as labeled data must be provided. Accordingly, configurations obtained from Monte Carlo simulations are grouped by noise value: those sampled below the critical noise are labeled ferromagnetic, while those above are labeled paramagnetic.

Additional preprocessing of the configurations is performed beyond noise separation. We did some \textit{data augmentation} by exploiting the $\mathbb{Z}_2$ symmetry of the MV model: for each configuration $\boldsymbol{s}$ generated by the simulation, its symmetric counterpart $\boldsymbol{s'} = -\boldsymbol{s}$ is also included. This procedure effectively doubles the available data.

\begin{figure}[ht!]
   \begin{center}
   \includegraphics[scale=0.4]{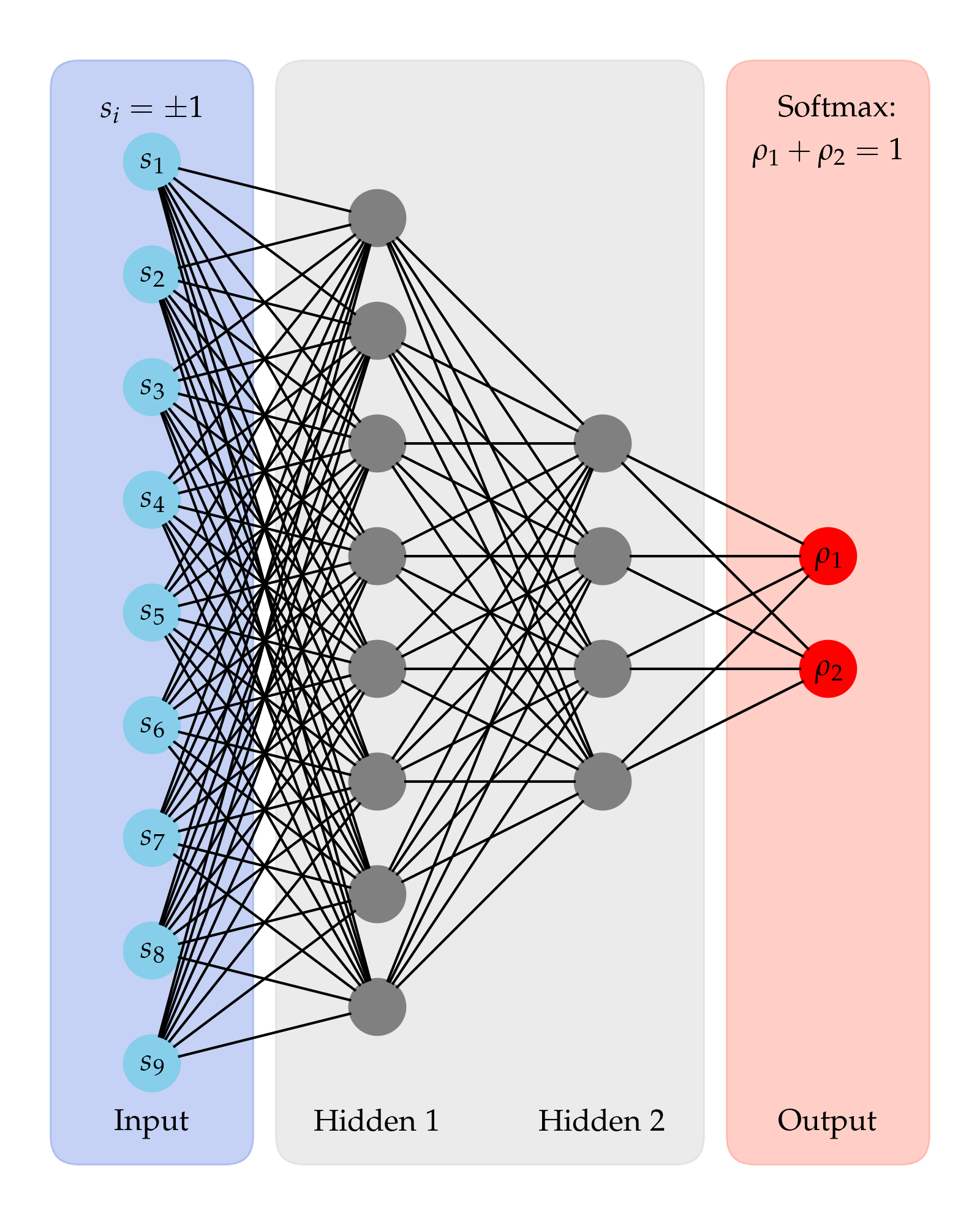}
   \end{center}
   \caption{Basic architecture of a neural network. Neurons, depicted as circles, are arranged in ordered layers from left to right. The input layer is on the left, the output layer on the right, and the intermediate layers are the hidden layers. Each neuron receives inputs from all neurons in the previous layer (dense connectivity), applies a linear transformation followed by a nonlinear activation function, and passes its output to the next layer. For spin configuration classification, the input layer consists of lattice spins $s_i = \pm 1$, and the output layer uses \textit{softmax} activation to produce normalized scores.}
   \label{rede-neural}
\end{figure}

The classification problem involves determining the degree to which a spin configuration resembles either a pure ferromagnetic state at $q=0$ or a paramagnetic state at $q_\infty$. For this purpose, a neural network classifier should employ a \textit{softmax} output layer with two neurons, ensuring the outputs are normalized, as illustrated in Fig.~(\ref{rede-neural}).

The neural network outputs two \textit{softmax} neurons, given by
\begin{equation}
   \rho_j = \varrho(g_j) = \frac{\exp\left(g_j\right)}{\exp\left(g_1\right) + \exp\left(g_2\right)}, \quad j = 1,2,
\end{equation}
where $\rho_j$ can be interpreted as the probability (or score) for each class label. For the MV model, each configuration yields outputs $(\rho_1, \rho_2)$, corresponding to the ferromagnetic and paramagnetic phases, respectively. In this context, the loss function can also be expressed as sparse categorical cross-entropy
\begin{equation}
   \ell_{\text{SCE}} = -\frac{1}{N_\mathcal{D}} \sum_{i=1}^{N_\mathcal{D}} y_i \ln y^\prime_i\left(\boldsymbol{\theta}\right),
   \label{loss-sce}
\end{equation}
where $\rho_{y_i}$ is the predicted probability for the true label $y_i$ of sample $i$.

The training data was obtained through Monte Carlo simulations, comprising $10^4$ configurations of the MV model for each of the 200 simulated noise values within the range $0.5 q^\square_c$ to $1.5 q^\square_c$ on the square lattice where $q^\square_c \sim 0.07518 $ is the critical noise of the MV model on the square lattice~\cite{Yu-2017}. After including the $\mathbb{Z}_2$ symmetry pairs for each configuration generated in the simulation, the total size of the training dataset reached $N_\mathcal{D} = 4 \times 10^6$ configurations. From this dataset, $20\%$ was reserved for validation. To shuffle the training data, simulation data, and their corresponding labels, the \textit{Scikit-Learn} library from the \textit{Python} programming language was utilized. 

During the Monte Carlo simulations, the initial $10^4$ steps were discarded to ensure that only stationary configurations were analyzed. To further reduce statistical correlations, an additional $10^3$ Monte Carlo steps were omitted between each stored configuration. Here, one Monte Carlo step corresponds to the update of all $N=L^2$ spins. The lattice sizes used throughout this study were $L=16$, $20$, $24$, $32$, and $40$. Configurations sampled at noise values below the critical noise ($q<q^\square_c$) were labeled as ferromagnetic, while those above ($q>q^\square_c$) were labeled as paramagnetic.

The neural network we used closely follows the dense architecture illustrated in Fig.~(\ref{rede-neural}), with the following structure:
\begin{itemize}
   \item An input layer of size $L^2$, where each input corresponds to a spin value $s_i = \pm 1$;
   \item A first hidden layer with 128 neurons, employing the \textit{ReLU} activation function and $\ell_2$ regularization, followed by batch normalization and a dropout layer with a rate of $0.2$;
   \item A second hidden layer with 64 neurons, also using \textit{ReLU} activation and $\ell_2$ regularization, followed by batch normalization and a dropout layer with a rate of $0.2$;
   \item An output layer with two neurons and \textit{softmax} activation.
\end{itemize}
The neural network was implemented and trained using the \textit{Keras} and \textit{Tensorflow} libraries in Python.

The output $\rho_1$ quantifies the similarity of the input configuration $\boldsymbol{s}$ to a pure ferromagnetic state ($q=0$), while $\rho_2$ quantifies similarity to a paramagnetic state ($q \to \infty$). Configuration labels are assigned as $y_i=1$ for the ferromagnetic phase and $y_i=0$ for the paramagnetic phase. The outputs are normalized, as depicted in Fig.~(\ref{rede-neural}), and the point of maximum confusion is reached when $\rho_1=\rho_2=0.5$. Training was performed for at least $10^3$ epochs with a batch size of $128$, using the \textit{ADAM} optimizer with a learning rate $\eta=10^{-4}$.

\begin{figure}[!ht]
   \begin{center}
   \includegraphics[scale=0.4]{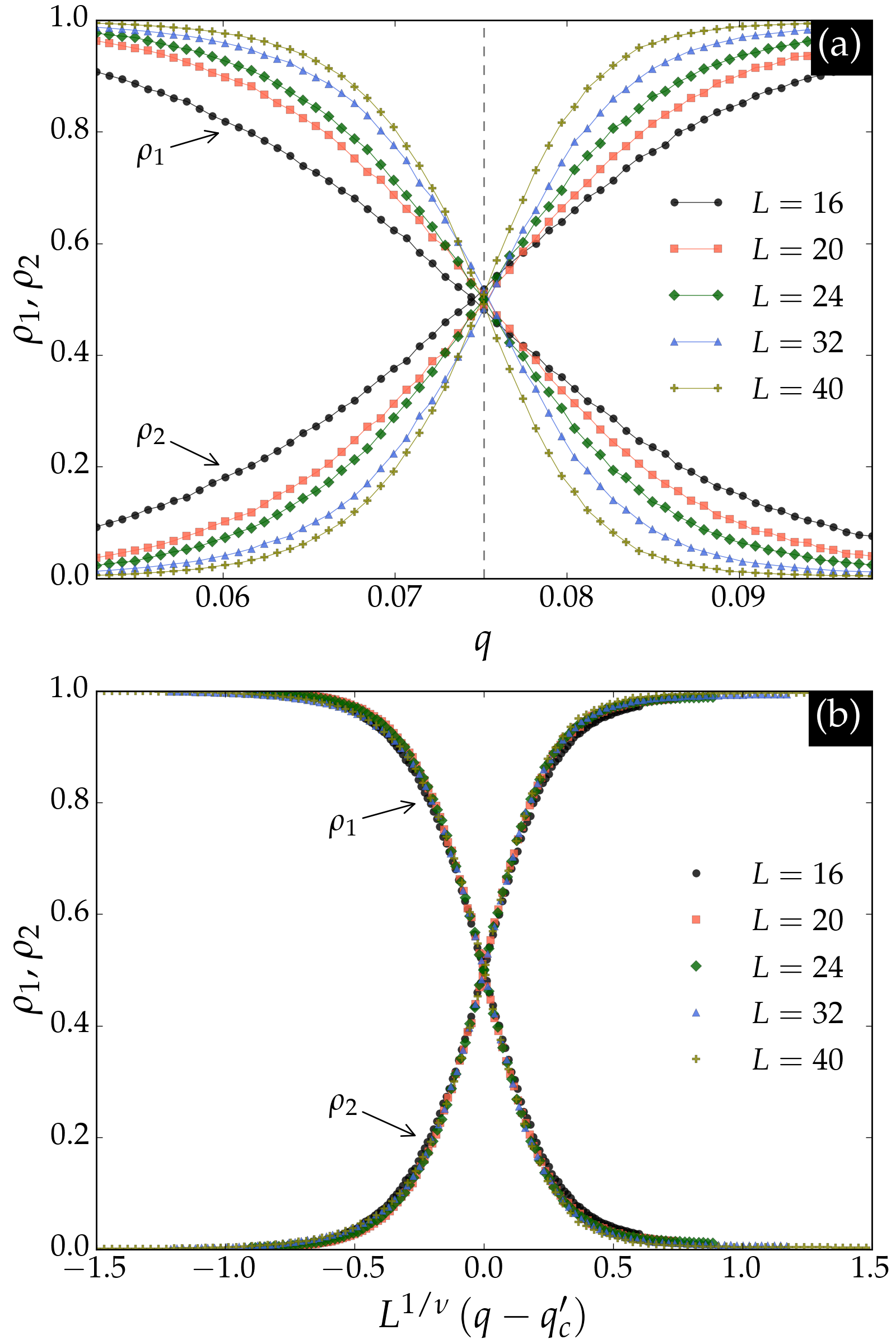}
   \end{center}
   \caption{Neural network output for the MV model on the square lattice, using inference data distinct from the training set. For each system size, two curves are shown: $\rho_1$ and $\rho_2$. The output $\rho_1$ (ferromagnetic phase) is close to $1$ at low noise values and decreases at high noise values, while $\rho_2$ (paramagnetic phase) behaves oppositely. The crossing of $\rho_1$ and $\rho_2$ marks the point of maximum confusion, where the network cannot distinguish between ferromagnetic and paramagnetic configurations. In panel (a), the crossing point $\rho_1=\rho_2=0.5$ closely matches the critical noise $q^\square_c$, indicated by the dashed vertical line. In panel (b), the outputs collapse according to Eq.~(\ref{classification-fss}) with the critical exponent $\nu=1$ for the square lattice; $q^\prime_c$ denotes the crossing abscissas.}
   \label{confidence-mv-square}
\end{figure}

For inference and model validation, we generated an additional $10^4$ configurations for $10^2$ noise values in the range $0.5 q^\square_c$ to $1.5 q^\square_c$, using Monte Carlo simulations with a different random seed. Results for the MV model on the square lattice are shown in Fig.~(\ref{confidence-mv-square}), using this inference dataset. The point of maximum confusion accurately identifies the critical point, and the neural network outputs obey the following scaling relations:
\begin{equation}
   \rho_{1,2} \propto f_{\rho_{1,2}}\left( N^{1/\nu} \left(q-q^\prime_{c}\right) \right), 
   \label{classification-fss}
\end{equation}
where $\nu$ is the correlation length exponent and $q^\prime_c$ is the crossing point of the output curves $\rho_1$ and $\rho_2$ as a function of noise. The functions $f_{\rho_1}$ and $f_{\rho_2}$ are scaling functions.

\begin{figure}[ht!]
   \begin{center}
   \includegraphics[scale=0.4]{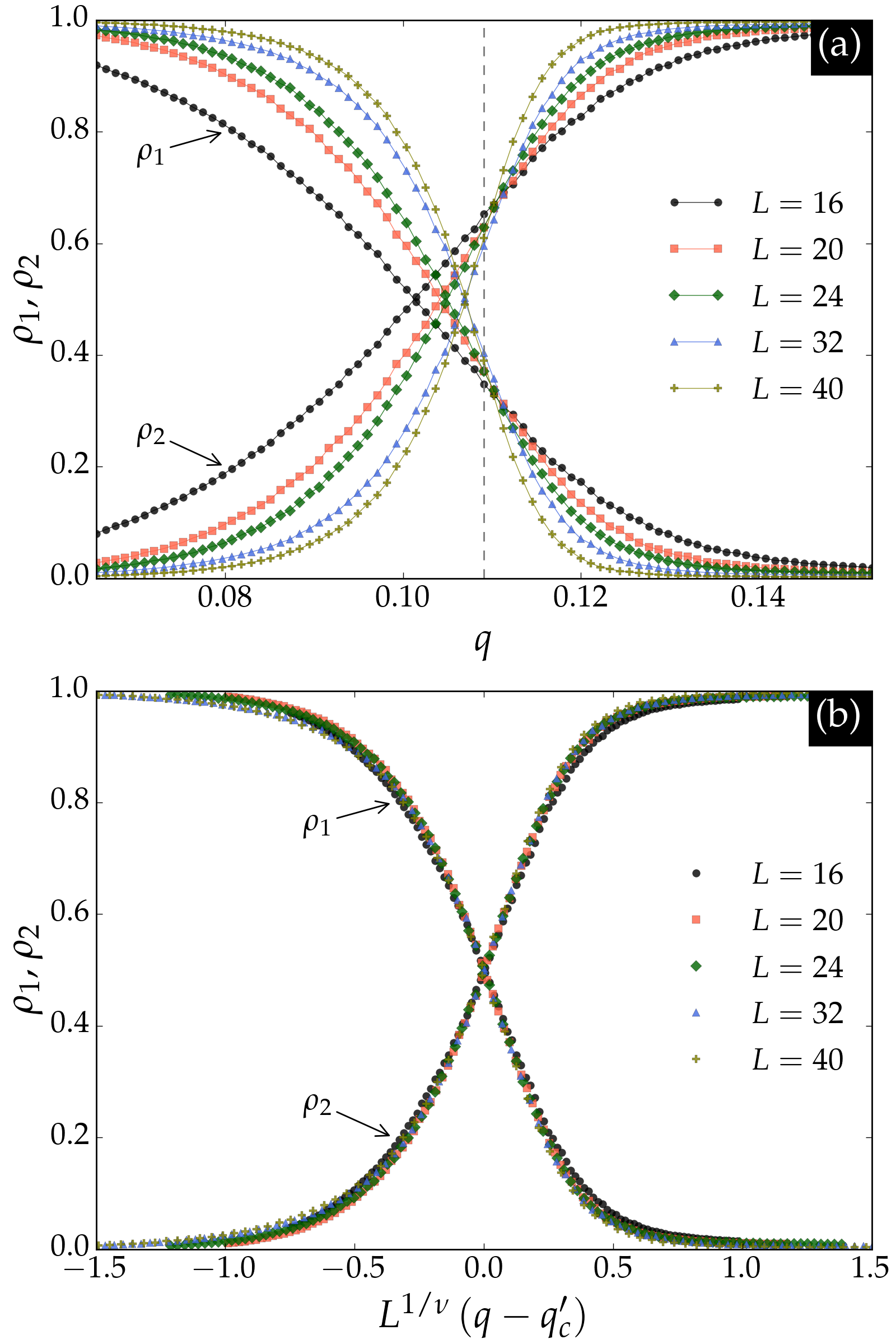}
   \end{center}
   \caption{Neural network outputs $\rho_1$ and $\rho_2$ for the MV model on the triangular lattice, trained with square lattice data. The curves have the same interpretation as in Fig.~(\ref{confidence-mv-square}). In panel (a), the crossing points $q^\prime_c$ ($\rho_1=\rho_2=0.5$) were used to estimate the critical noise via the process in Fig.~(\ref{regression-mv-triangular}). The critical noise $q^\triangle_c$ is indicated by the dashed vertical line. In panel (b), the outputs scale according to Eq.~(\ref{classification-fss}) with critical exponent $\nu=1$.}
   \label{confidence-mv-triangular}
\end{figure}

Similarly, we generated inference data for the MV model on the triangular lattice: $10^4$ configurations for $10^2$ noise values in the range $0.5 q^\triangle_c$ to $1.5 q^\triangle_c$, where $q^\triangle_c \sim 0.10910$ is now the critical noise for the triangular lattice~\cite{Yu-2017}. We show a triangular lattice in panel (b) of Fig.~(\ref{lattices}). The neural network outputs for the triangular lattice are presented in Fig.~(\ref{confidence-mv-triangular}). In panel (a), the crossing point $q^\prime_c$ approaches the critical noise $q^\triangle_c$ as the system size increases. The scaling behavior of the outputs matches that of the square lattice, with curves collapsing according to Eq.~(\ref{classification-fss}) and critical exponent $\nu=1$. The only difference between the square and triangular lattices is the non-universal noise $q_c$. Models trained on square lattice data can be used to estimate critical points for other lattices, provided the number of spins is the same.

\begin{figure}[ht!]
   \begin{center}
   \includegraphics[scale=0.45]{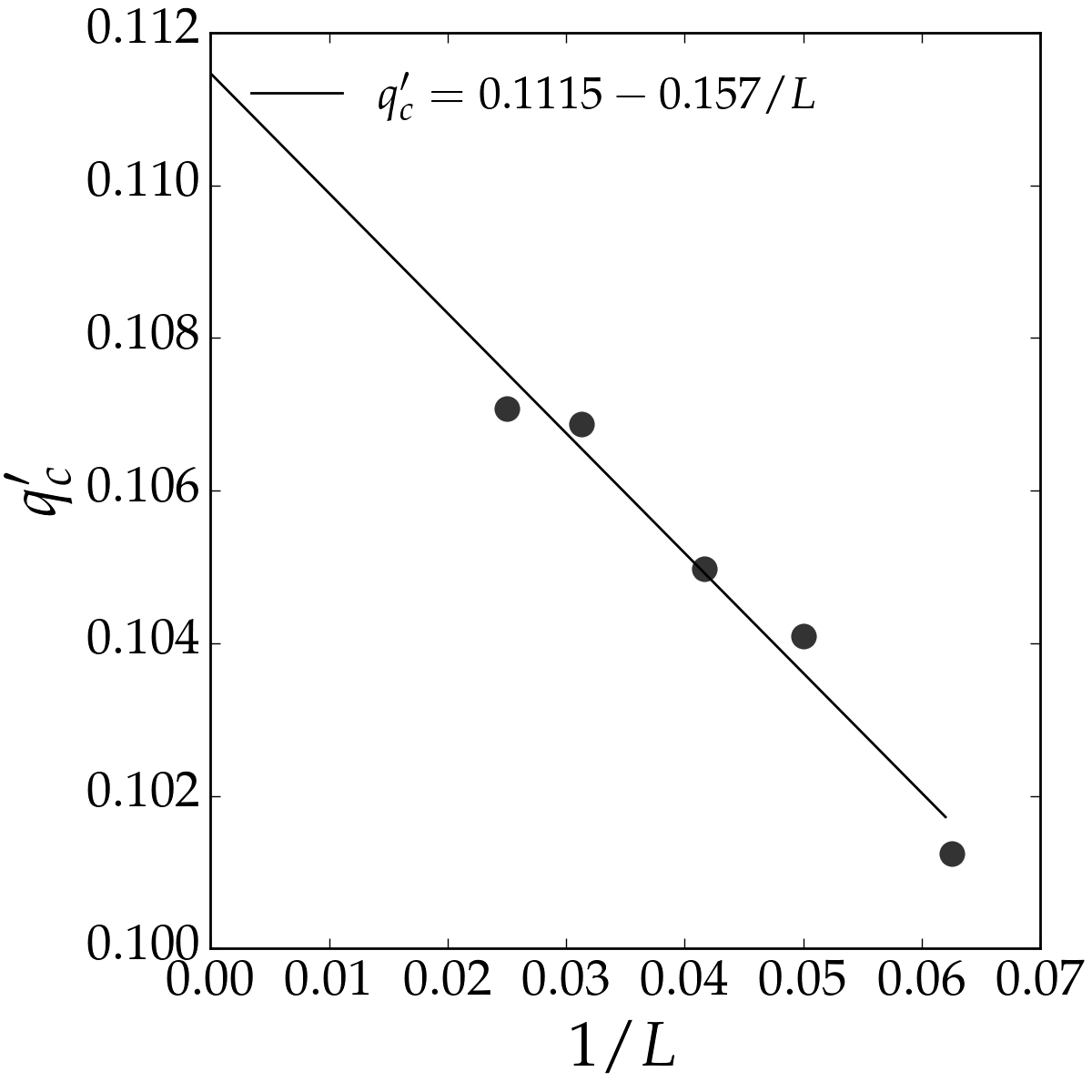}
   \end{center}
   \caption{Linear regression of the crossing points $q^{\prime}_c$ of the neural network outputs $\rho_1$ and $\rho_2$ for MV model configurations on the triangular lattice. Extrapolation according to Eq.~(\ref{regression-threshold}) yields an estimate for the critical noise, $q_c \sim 0.1115 \pm 0.0007$, which is close to the critical noise $q^\triangle_c \sim 0.10910$ for the MV model on the triangular lattice~\cite{Yu-2017}, obtained via Monte Carlo simulations.}
   \label{regression-mv-triangular}
\end{figure}

Using the model trained on square lattice data, we inferred the critical noise for the triangular lattice by collecting the crossing points $q^\prime_c$ of $\rho_1$ and $\rho_2$ and performing a linear regression
\begin{equation}
   q^{\prime}_c = q_c - a/L,
   \label{regression-threshold}
\end{equation}
yielding $q_c \sim 0.1115 \pm 0.0007$, which is very close to the critical noise $q^\triangle_c \sim 0.10910$ of the triangular lattice~\cite{Yu-2017}. The regression results are shown in Fig.~(\ref{regression-mv-triangular}).

We can therefore conclude the general applicability of the classification neural network method for estimating the critical thresholds of stochastic models. This approach is effective even for irreversible models exhibiting nonequilibrium critical phenomena. Moreover, models trained on data from one lattice can be used to estimate the critical points of other lattices, provided they have the same number of spins.

\section{Unsupervised Learning}

We now present the results obtained from unsupervised learning applied to the MV model. Two unsupervised techniques were employed: \textit{PCA} and variational autoencoder (\textit{VAE}). \textit{VAEs} are a special class of autoencoders, which are generative models capable of producing new data instances that closely resemble the training data. While applications of these methods to the Ising~\cite{Wang-2016, Hu-2017, Wetzel-2017, Mehta-2019, Walker-2020} and Potts~\cite{Tirelli-2022} models have been previously reported, here we provide original results for the MV model, demonstrating how these techniques can be used to estimate the critical threshold of the continuous phase transition. Both approaches enable characterization of the phase transition and estimation of the critical point and critical exponents in the MV model.

\subsection{PCA}

The foundation of the \textit{PCA} technique is to first identify the hyperplane that captures the greatest variance in the data and then project the data onto this hyperplane. The first principal component is the direction that preserves as much of the data variance as possible. The second principal component captures the largest remaining variance, and so forth.

The directions of the $d$-dimensional hyperplane are called principal components, and they are determined via a linear transformation. Explicitly, the principal components are the eigenvectors of the covariance matrix. One can interpret the \textit{PCA} as fitting the dataset inside a $d$-dimensional ellipsoid, where the semi-axes correspond to the variances of the principal components, ordered from largest to smallest. Thus, the \textit{PCA} technique assumes the data is centered at zero, which naturally occurs for spin configurations when their $\mathbb{Z}_2$ symmetric pairs are included.

For multidimensional data such as spin configurations, the first few principal components contain most of the relevant information about the data. This forms the basis of dimensionality reduction. Dimensionality reduction also results in feature selection, which leads to one of the main applications of \textit{PCA}: data compression while preserving key characteristics such as variance. The inverse linear transformation of the \textit{PCA} acts as decompression, restoring the original dimensionality of the data~\cite{Geron-2019}.

We applied \textit{PCA} to the same training data for the MV model described in the previous section. The implementation utilized the \textit{Scikit-Learn} library in Python, which computes the eigenvalues and eigenvectors of the covariance matrix. In \textit{PCA}, the eigenvectors (principal components $\boldsymbol{p}_i$) represent directions in the data space along which the variance is maximized, ordered from largest to smallest. The corresponding eigenvalues $\lambda_i$ quantify the variance along each principal component. The number of principal components equals the number of spin configurations in the dataset.

\begin{figure}[ht!]
   \begin{center}
   \includegraphics[scale=0.4]{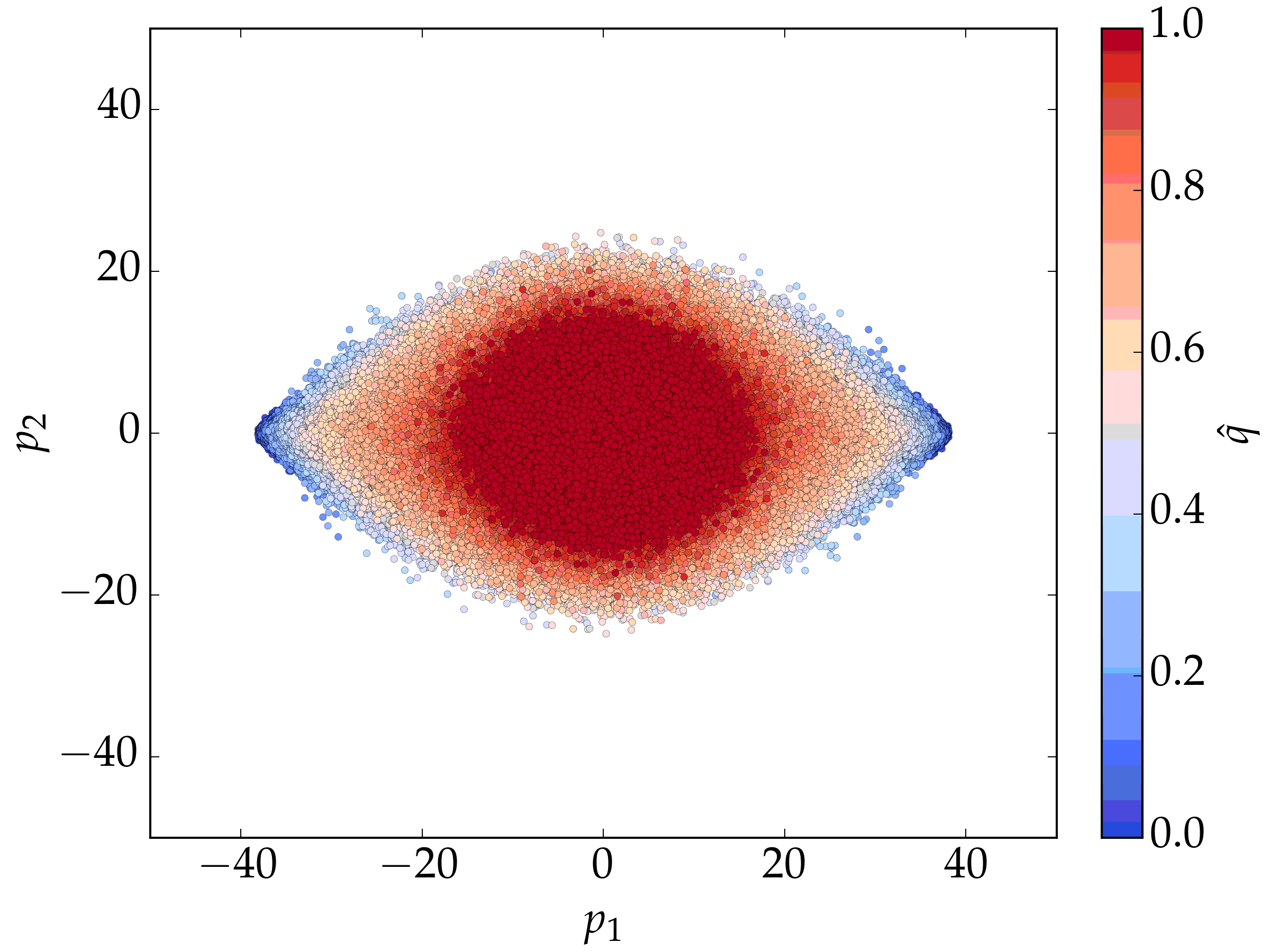}
   \end{center}
   \caption{Projection of the MV model training data in a square lattice of $L=40$ onto the first two principal components as a function of the noise parameter. Here, \textit{PCA} was applied separately for each noise value in the training set, and $\hat{q}$ denotes normalized noise values from $0.5q^\square_c$ to $1.5q^\square_c$. For the lowest noise values, the principal components form two clusters centered at $(0,-L)$ and $(0,L)$; for higher noise values, a single cluster appears at $(0,0)$. This clustering enables visualization of the phase transition, as reflected in the principal component data.}
   \label{pca-components}
\end{figure}

Fig.~(\ref{pca-components}) displays the projection of the data onto the first two principal components, $p_1$ and $p_2$, for the training set. Points are colored according to the noise parameter, ranging from $0.5q^\square_c$ to $1.5q^\square_c$. At low noise values, two distinct clusters emerge, centered at $(0,-L)$ and $(0,L)$, while at high noise values, a single cluster appears at $(0,0)$. This clustering structure reveals the presence of a phase transition: at low noise, the histogram of the principal components exhibits two separated maxima, whereas at high noise, only one maximum is present. The transition between these regimes occurs as the maxima merge at an intermediate noise value.

The most important observables for \textit{PCA} behave near the critical noise according to the following scaling relations:
\begin{eqnarray}
   \lambda_2/\lambda_1 &\propto& f_{\lambda}\left( N^{1/\nu} \left(q-q_{c}\right) \right), \nonumber \\
   P_1/L  &\propto& N^{-\beta/\nu} f_{P_1}\left( N^{1/\nu} \left(q-q_{c}\right) \right), \nonumber \\
   L P_2  &\propto& N^{\gamma/\nu} f_{P_2}\left( N^{1/\nu} \left(q-q_{c}\right) \right), 
   \label{observables-pca-fss}
\end{eqnarray}
where $\lambda_2/\lambda_1$ is the ratio between the eigenvalues of the respectively first two principal components. Equivalently, $\lambda_2/\lambda_1$ is also the ratio of the second to the first principal component explained variance. Furthermore,
\begin{equation}
      P_{1,2} = \left\langle \left| p_{1,2} \right| \right\rangle,
\end{equation}
are the averages of the absolute values of the projections of the data onto the first and second principal components, respectively.

\begin{figure*}[p]
   \begin{center}
   \includegraphics[scale=0.4]{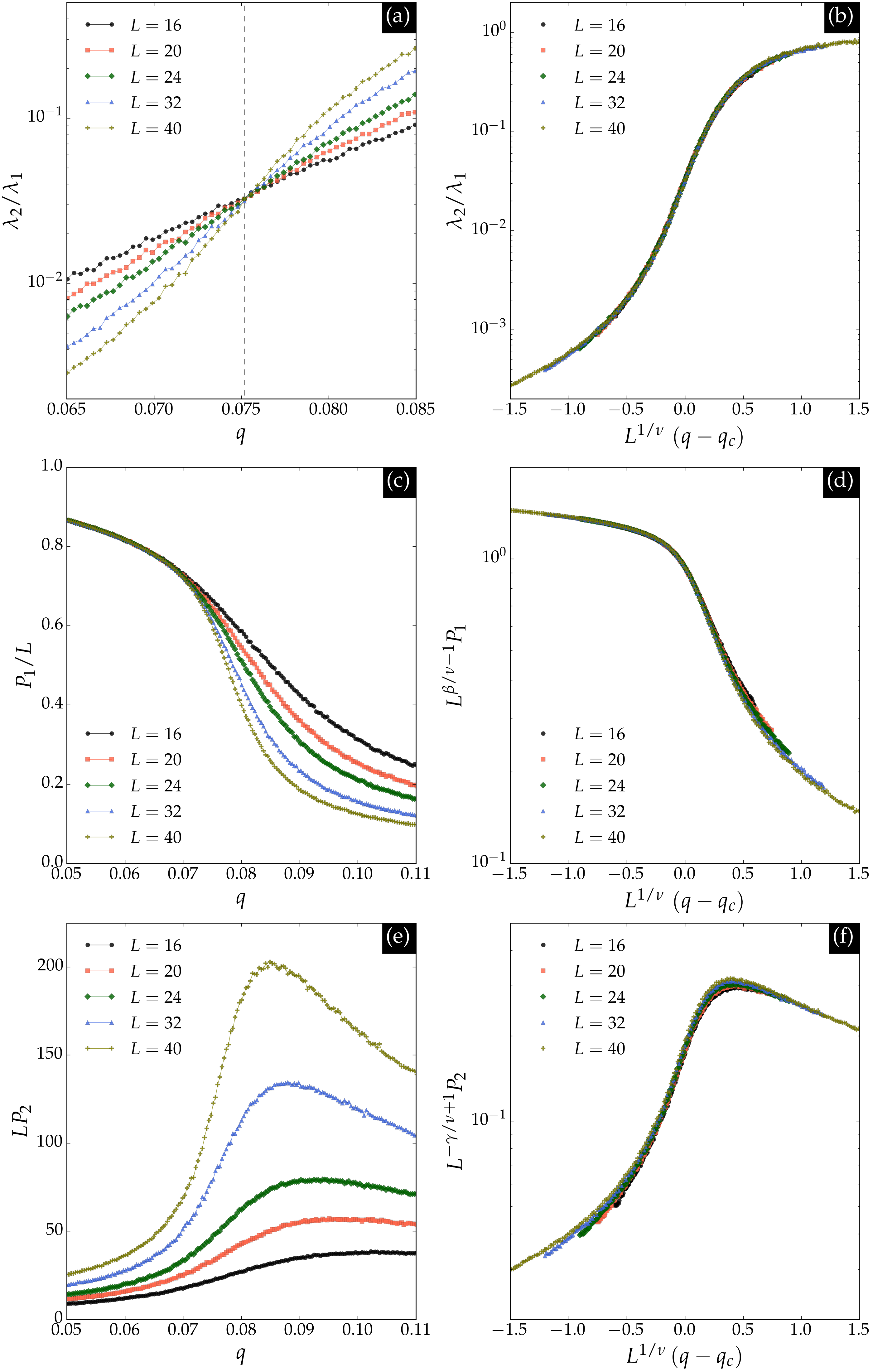}
   \end{center}
   \caption{Characterization of the MV model phase transition by the ratio of the two largest eigenvalues $\lambda_2/\lambda_1$ and by the averages of the absolute values of the principal components $P_1$ and $P_2$. In panel (a), we show the ratio between the two largest eigenvalues $\lambda_2/\lambda_1$. We note that $\lambda_2/\lambda_1$ is universal at the critical noise $q^\square_c$ of the square lattice. In panel (b), we observe that the scaling transformation allows us to estimate the critical exponent $\nu=1$. In panel (c), we have $P_1/L$ as a function of noise, which coincides with the average magnetization per spin. The magnetization scales with $L^{\beta/\nu}$ where $\beta/\nu=1/8$, as shown in panel (d). In panel (e), we show $LP_2$, whose maximum increases with $L^{\gamma/\nu}$ at the critical point where $\gamma/\nu=7/4$, as shown in panel (f).}
   \label{pca-fss}
\end{figure*}

According to Eq.~(\ref{observables-pca-fss}), the ratio $\lambda_2/\lambda_1$ is universal for the MV model. Therefore, the curves of the ratio $\lambda_2/\lambda_1$ should cross at the critical noise of the MV model. The ratio $\lambda_2/\lambda_1$ plays a role analogous to the Binder cumulant in Monte Carlo simulations, being universal at the critical noise and allowing its estimation by locating the crossing point of the curves for different system sizes $L$. Results for the MV model are shown in panel (a) of Fig.~(\ref{pca-fss}). The data collapse in panel (b) confirms the scaling relation for $\lambda_2/\lambda_1$ and allows estimation of the critical exponent $\nu=1$.

We also identify $P_1/L$, shown in panel (c) of Fig.~(\ref{pca-fss}), as the average magnetization per spin. Since the magnetization per spin is expected to scale with $L^{\beta/\nu}$, the data from the first principal component enable us to estimate the exponent $\beta=1/8$, as shown in the collapse in panel (d). Although there is no \textit{a priori} guarantee that the projections onto the first principal component correspond to the magnetization, this correspondence can be interpreted through the fluctuation-dissipation theorem. According to this theorem, the fluctuation of the magnetization is proportional to the magnetic susceptibility, which diverges at the critical noise. Since the variance associated with the magnetization diverges at the critical point, it can be expected for the first principal component to be related to the magnetization.

Furthermore, the mean $P_2 L$, shown in panel (e) of Fig.~(\ref{pca-fss}), is proportional to the variance of the magnetization. Again, by the fluctuation-dissipation theorem, the exponent $\gamma=7/4$ can be estimated from the data of the second principal component, as shown in panel (f), whose maximum should scale as $L^{\gamma/\nu}$. Thus, the \textit{PCA} technique allows us to estimate three critical exponents, which is sufficient to determine the universality class of the two-dimensional Ising model, since the remaining exponents can be obtained via scaling relations.

\subsection{VAEs}

Autoencoders are a class of neural network architectures designed to learn compact representations of data in a lower-dimensional vector space, known as the latent space $\boldsymbol{z}$. This latent space encodes the essential features of the input data, typically with far fewer dimensions than the original input, enabling both dimensionality reduction and feature selection. The mapping from input to latent space is performed in an unsupervised fashion, allowing the autoencoder to capture the most relevant characteristics of the data without requiring labeled examples.

Another prominent generative model architecture alongside \textit{VAEs} are the generative adversarial networks (\textit{GANs}), which can also generate realistic samples from the input data distribution. Both \textit{VAEs} and \textit{GANs} aim to learn a meaningful latent representation of the data, but they differ significantly in their training procedures. \textit{VAEs} are trained by minimizing a reconstruction loss, encouraging the model to accurately reproduce the input at the output, while also regularizing the latent space. In contrast, \textit{GANs} employ a game-theoretic approach involving two neural networks: a generator, which creates synthetic data, and a discriminator, which attempts to distinguish between real and generated samples.

Here, we consider the \textit{VAEs}, which are employed to detect the ferromagnetic-paramagnetic phase transition of the MV model in an unsupervised manner. As we will show, the loss function associated with the \textit{VAE} can be used to identify the symmetry breaking associated with the transition of MV model and related spin systems.

The theoretical foundation of the latent space $\boldsymbol{z}$ mapping in \textit{VAEs} is rooted in Bayesian inference theory~\cite{deGroot-2012}, which finds applications in fields such as forensic science and data analysis. \textit{VAEs} implement variational Bayesian statistics, where the objective is to maximize the marginal likelihood $p\left( \mathcal{D} \right)$ of the dataset $\mathcal{D}$ with respect to the model hyperparameters $\boldsymbol{\theta}$. This approach is known as maximum likelihood estimation.

In Bayesian inference, we consider a model characterized by latent variables $\boldsymbol{z}$, described by the posterior distribution $p\left(\boldsymbol{z} \mid \mathcal{D}\right)$. The posterior represents the probability of the latent variables $\boldsymbol{z}$ given the observed dataset $\mathcal{D}$, and is computed via Bayes theorem:
\begin{equation}
   p\left(\boldsymbol{z} \mid \mathcal{D}\right) = \frac{p\left(\mathcal{D} \mid \boldsymbol{z}\right) p\left(\boldsymbol{z}\right)}{p\left(\mathcal{D}\right)},
   \label{Bayes-theorem}
\end{equation}
where $p\left(\mathcal{D} \mid \boldsymbol{z}\right)$ is the likelihood, $p\left(\boldsymbol{z}\right)$ is the prior, which is often chosen as a multivariate normal distribution $\mathcal{N}(\boldsymbol{z} \vert \boldsymbol{0},\boldsymbol{I})$. The denominator $p\left(\mathcal{D}\right)$ is the marginal likelihood, given by
\begin{equation}
   p\left( \mathcal{D} \right) = \int p\left(\mathcal{D} \mid \boldsymbol{z}\right) p\left(\boldsymbol{z}\right) \,d\boldsymbol{z}.
   \label{marginal-likelihood}
\end{equation}
There is a close analogy between the marginal likelihood and the partition function in statistical mechanics, with $p\left(\mathcal{D} \mid \boldsymbol{z}\right) p\left(\boldsymbol{z}\right)$ playing the role of the Boltzmann weight.

The central idea of variational Bayesian inference is to approximate the true posterior $p_\theta(\boldsymbol{z} \vert \mathcal{D})$ by a simpler distribution $q_\theta(\boldsymbol{z})$, called the variational distribution, and to make this variational distribution as close as possible to the prior $p_\theta(\boldsymbol{z})$ while still allowing accurate reconstruction of the data. In the context of VAEs, this is achieved by minimizing the Kullback-Leibler (KL) divergence between $q_\theta(\boldsymbol{z})$ and $p_\theta(\boldsymbol{z})$, which regularizes the latent space and encourages $q_\theta(\boldsymbol{z})$ to remain close to the prior.

Explicitly, the variational distribution is often chosen as
\begin{equation}
   q_{\theta}\left(\boldsymbol{z} \right) = \mathcal{N} \left(\boldsymbol{z} \vert \boldsymbol{\mu}, \boldsymbol{\sigma}^2 \right),
   \label{distro-vae}
\end{equation}
where $\boldsymbol{\mu}$ and $\boldsymbol{\sigma}$ are the mean and standard deviations, respectively, which can be learned by a neural network (the encoder). One can use the following reparameterization
\begin{equation}
   \boldsymbol{z} = \boldsymbol{\mu} + \boldsymbol{\sigma} \odot \boldsymbol{\epsilon}, \quad \text{with } \boldsymbol{\epsilon} \sim \mathcal{N}(\boldsymbol{0}, \boldsymbol{I}),
\end{equation}
which allows the use of the backpropagation algorithm. This technique is known as the ``reparameterization trick''.

Finally, the problem of Bayesian inference is transformed into the minimization of the total loss function of the \textit{VAE}, $\ell_\text{VAE}$, given by
\begin{equation}
\ell_\text{VAE} = \ell_\text{RECON} + \ell_\text{KL},
\end{equation}
where $\ell_\text{RECON}$ is the reconstruction loss and $\ell_\text{KL}$ is the regularization term. The reconstruction loss $\ell_\text{RECON}$ measures how well the decoder can reconstruct the input data from the latent representation, while the regularization term $\ell_\text{KL}$ ensures that the learned variational distribution $q_{\theta}\left(\boldsymbol{z} \right)$ remains close to the prior distribution $p_\theta(\boldsymbol{z})$. 

The balance between the reconstruction and regularization losses allows the model to learn meaningful features while maintaining a structured latent space. The $\ell_\text{KL}$ is the Kullback-Leibler loss function~\cite{Kullback-1951} between the variational distribution and the prior $p_\theta(\boldsymbol{z})$. In the case of both prior and variational distributions given by normal distributions, the loss function $\ell_\text{KL}$ can be written as
\begin{equation}
   \ell_\text{KL} = - \frac{1}{2} \sum_{i=1}^{d} \left(1 + \log \sigma_i^2 - \mu_i^2 - \sigma_i^2\right).
   \label{kl-loss}
\end{equation}
and the reconstruction loss function $\ell_\text{RECON}$ can be given by the mean squared error loss function $\ell_\text{MSE}$, given by Eq.~(\ref{loss-mse}), or by the binary cross-entropy loss function $\ell_\text{BCE}$
\begin{equation}
  \ell_{\text{BCE}} = - \sum^{N_{\mathcal{D}}}_{i=1} y_i \ln y^\prime_i(\boldsymbol{\theta}) - (1 - y_i) \ln \left[ 1 - y^\prime_i(\boldsymbol{\theta}) \right].
  \label{loss-bce}
\end{equation}

\begin{figure*}[htb!]
   \begin{center}
   \includegraphics[scale=0.4]{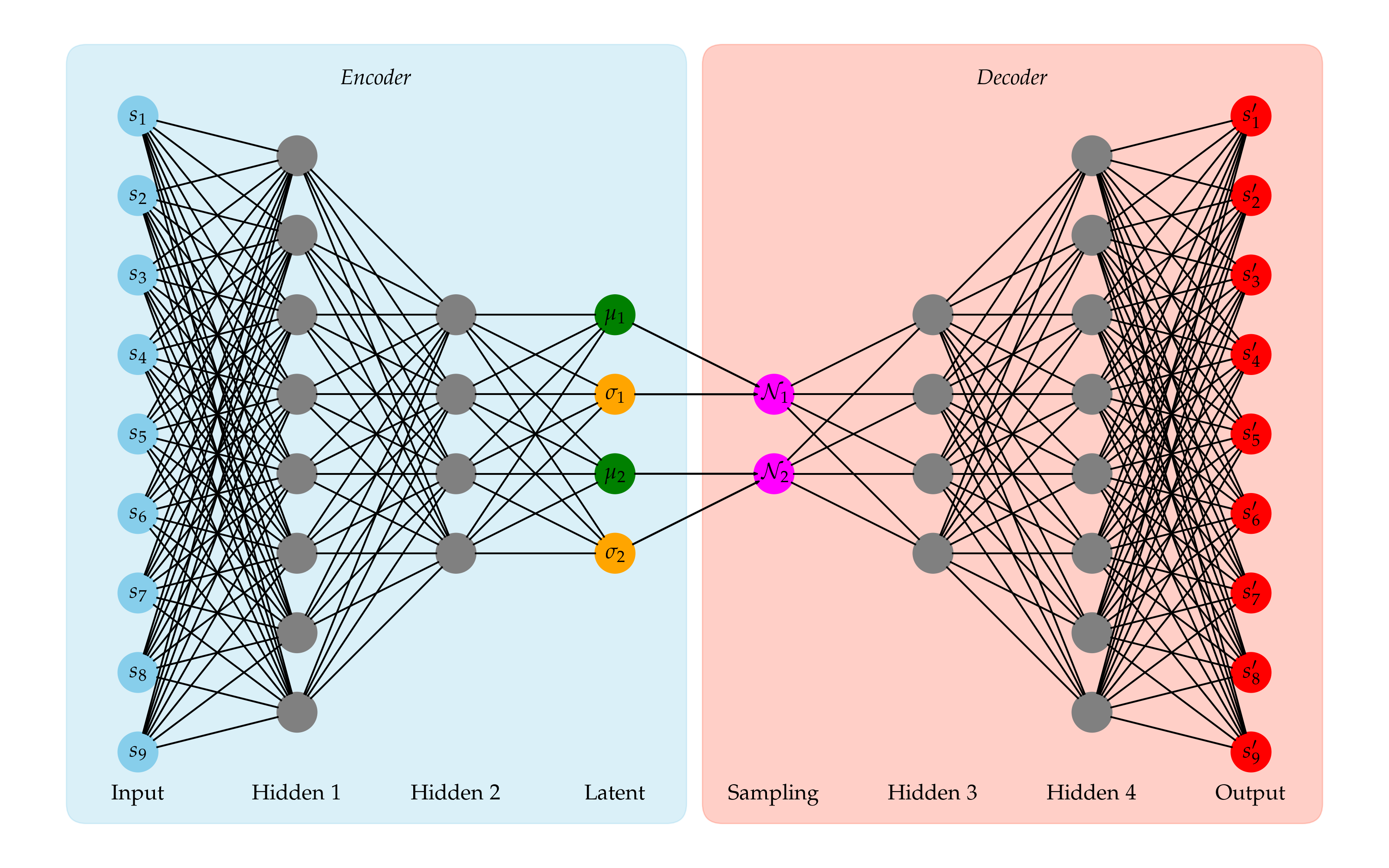}
   \end{center}
   \caption{Basic architecture of \textit{VAEs}. The key difference from a standard autoencoder is that the encoder outputs the means $\boldsymbol{\mu}$ and standard deviations $\boldsymbol{\sigma}$, which are passed to a custom sampling layer. The actual latent variables $z_i$ are obtained by sampling from normal distributions $\mathcal{N}_i(\mu_i, \sigma_i)$, which are then input to the decoder. Because the latent representation is a statistical distribution, \textit{VAEs} are generative models: new artificial data can be generated by sampling from the prior distribution $p_\theta(\boldsymbol{z})$ and decoding the result.}
   \label{variational-autoencoder}
\end{figure*}

Fig.~(\ref{variational-autoencoder}) illustrates the basic architecture of a \textit{VAE}. The encoder outputs the means $\boldsymbol{\mu}$ and standard deviations $\boldsymbol{\sigma}$ of the latent variables, which are then passed to a sampling layer. This custom layer produces latent variables $z_i$ by sampling from the distributions $\mathcal{N}_i(\mu_i, \sigma_i)$. The \textit{VAE} is also a generative model: new artificial data instances can be generated by sampling from the prior distribution $p_\theta(\boldsymbol{z})$ and decoding the result.

To obtain the results in this section, we employed \textit{VAEs} with an architecture similar to that shown in Fig.~(\ref{variational-autoencoder}). The encoder consists of the following layers:
\begin{itemize}
   \item An input layer of size $L^2$, where each input corresponds to a spin value $s_i = \pm 1$ from the configuration;
   \item A first hidden layer with 625 neurons, using the \textit{ReLU} activation function and $\ell_1$ regularization to promote feature selectivity, followed by batch normalization and a dropout layer with a rate of $0.2$;
   \item A second hidden layer with 256 neurons, also with \textit{ReLU} activation and $\ell_1$ regularization, followed by batch normalization and dropout ($0.2$);
   \item A third hidden layer with 64 neurons, again with \textit{ReLU} activation, $\ell_1$ regularization, batch normalization, and dropout ($0.2$);
   \item An output layer with two neurons with linear activation: one outputs the mean $\mu$ of the latent variable $Z$ (sampled from a normal distribution), and the other outputs the logarithm of the variance $\sigma$ of $Z$.
\end{itemize}
The decoder, which receives the latent encoding $Z$ as input, mirrors the encoder's structure. In addition, the decoder includes an extra input neuron for the normalized noise of the configuration, making the architecture a conditional \textit{VAE}.

The \textit{VAE} was trained using the same dataset described in the previous section for the MV model, consisting of $10^4$ configurations for each of $10^2$ noise values in the range from $0.5 q^\square_c$ to $1.5 q^\square_c$. Training was performed using a combined loss function comprising the mean squared error (MSE) and the Kullback-Leibler (KL) divergence, as given in Eq.~(\ref{kl-loss}), with the \textit{RMSprop} optimizer. Accordingly, the output layer neurons employed hyperbolic tangent (\textit{tanh}) activation. When the binary cross-entropy loss $\ell_\text{BCE}$ and the \textit{sigmoid} activation for the output neurons were used instead of $\ell_\text{MSE}$ and \textit{tanh} activation, the results remained qualitatively unchanged. We used a learning rate $\eta=10^{-3}$ and a batch size of $128$. The model was trained for $10^3$ epochs.

\begin{figure*}[htb!]
   \begin{center}
   \includegraphics[scale=0.35]{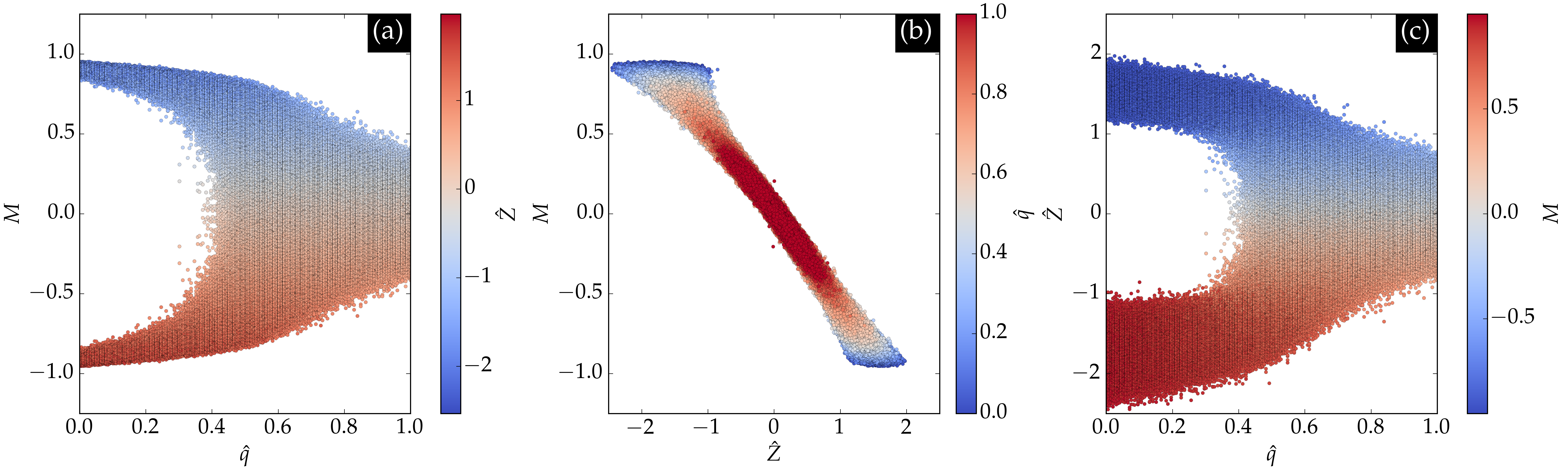}
   \end{center}
   \caption{Dependence of the normalized latent encoding $\hat{Z}$ on magnetization and noise for the MV model data. Panel (a) shows the magnetizations of input configurations as a function of normalized noise from $0.5q^\square_c$ to $1.5q^\square_c$, with the color gradient representing the latent encoding $Z$. Both magnetizations and encodings respect the $\mathbb{Z}_2$ inversion symmetry. Panel (b) displays the nearly linear relationship between magnetization and latent encoding. Panel (c) illustrates that the phase transition is also evident from the latent encoding data.}
   \label{vae-latent}
\end{figure*}

Fig.~(\ref{vae-latent}) shows the latent encoding $Z$ of the input data as a function of magnetization and normalized noise. In panel (a), we observe a clear separation between positive and negative magnetizations according to the sign of $Z$, reflecting the $\mathbb{Z}_2$ symmetry. Panel (b) demonstrates that the relationship learned by the neural network between magnetization $m$ and latent encoding $Z$ is approximately linear. Notably, two clusters centered at $(-2,1)$ and $(2,-1)$ appear at low noise values, while a single cluster at $(0,0)$ emerges at higher noise values, indicating the phase transition. Panel (c) further confirms that the phase transition can be detected from the latent encoding.

We can determine the critical point using the correlation function between the real data configurations $\boldsymbol{s}_\text{real}$ and those reconstructed by the \textit{VAE} $ \boldsymbol{s}_\text{recon}$, defined as
\begin{equation}
   C \left(\boldsymbol{s}_\text{real} \mid \boldsymbol{s}_\text{recon} \right) \equiv \frac{1}{L^2} \frac{\left\langle \left| \boldsymbol{s}_\text{real} \cdot \boldsymbol{s}_\text{recon} \right| \right\rangle}{m_\text{real} m_\text{recon} },
   \label{autoencoder-correlation}
\end{equation}
where $\langle \cdots \rangle$ denotes an average over the dataset, $m_\text{real}$ is the average magnetization of the real data, and $m_\text{recon}$ is the average magnetization of the reconstructed data. The correlation function is universal at the critical point, enabling estimation of the transition threshold in the same way as the Binder cumulant in Monte Carlo simulations
\begin{equation}
   C \left(\boldsymbol{s}_\text{real} \mid \boldsymbol{s}_\text{recon} \right) \propto f_{C}\left( N^{1/\nu} \left(q-q_{c}\right) \right),
   \label{correlation-vae-fss}
\end{equation}
which allows the estimation of the correlation length exponent $\nu$.

We can also analyze the loss functions $\ell_\text{MSE}$ and $\ell_\text{BCE}$ between the input and reconstructed output configurations, which also serve as indicators of the phase transition. In the paramagnetic regime ($T \rightarrow \infty$), the input and reconstructed outputs behave as two effectively random configurations, yielding limiting values $\ell_\text{MSE} \rightarrow 1$ and $\ell_\text{BCE} \rightarrow \ln{2}$ for random binary data. Consequently, the quantities $1-\ell_\text{MSE}$ and $1-\ell_\text{BCE}/\ln 2$ act as order parameters and obey the following scaling relations:
\begin{eqnarray}
   1-\ell_\text{MSE} &\propto& L^{2\beta/\nu} f_\text{MSE}\left( N^{1/\nu} (q-q_{c}) \right), \nonumber \\
   1-\ell_\text{BCE}/\ln{2} &\propto& L^{2\beta/\nu} f_\text{BCE}\left( N^{1/\nu} (q-q_{c}) \right).
   \label{losses-vae-fss}
\end{eqnarray}
Since the loss functions depend on both the input and output configurations, and the average magnetizations of each scale with exponent $\beta/\nu = 1/8$, it follows that the order parameters above should scale with system size as $2\beta/\nu = 1/4$, consistent with the universality class of the two-dimensional Ising model.

\begin{figure*}[p]
   \begin{center}
   \includegraphics[scale=0.4]{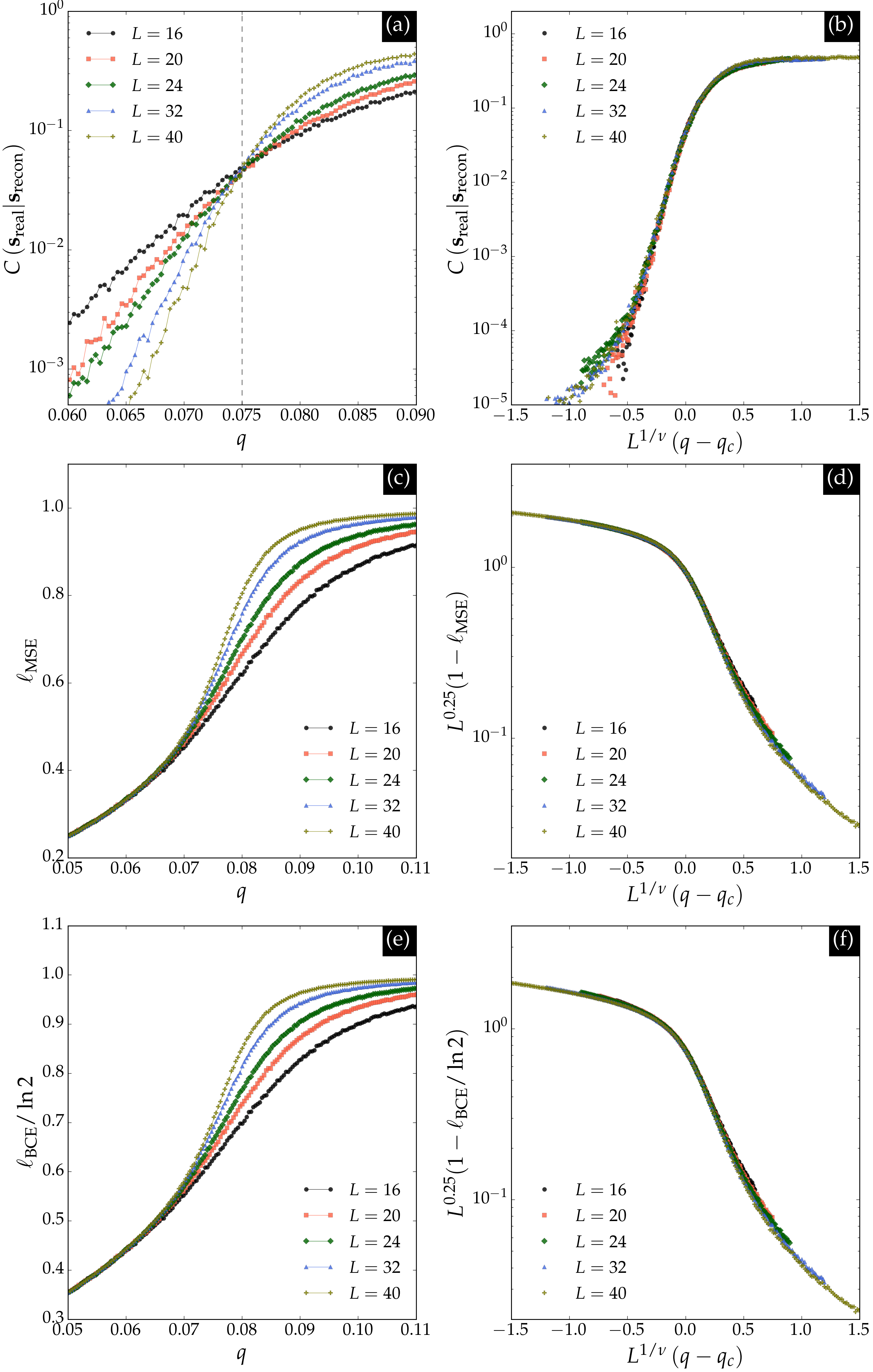}
   \end{center}
   \caption{Correlations and loss functions between the input configurations of the MV model and the reconstructed outputs from the \textit{VAE}. The correlation function, computed as in Eq.~(\ref{autoencoder-correlation}), is shown in panel (a). At the critical noise $q^\square_c$ (vertical dashed line), the correlation function is universal. The scaling collapse in panel (b) confirms the relation in Eq.~(\ref{correlation-vae-fss}) with the Ising exponent $\nu=1$. Panels (c) and (e) show that $1-\ell_\text{MSE}$ and $1-\ell_\text{BCE}/\ln 2$ serve as order parameters, vanishing at the transition. Their scaling collapses in panels (d) and (f) follow Eq.~(\ref{losses-vae-fss}) with exponent $2\beta/\nu = 1/4$.}
   \label{vae-fss}
\end{figure*}

Results for the MV model are presented in Fig.~(\ref{vae-fss}). Panel (a) shows the correlation function between the input and reconstructed configurations, which is universal at the critical noise $q^\square_c$ of the square lattice. The scaling collapse in panel (b) confirms the scaling relation for the correlation function and allows estimation of the critical exponent $\nu=1$. Panels (c) and (e) display $\ell_\text{MSE}$ and $\ell_\text{BCE}/\ln 2$, respectively, both of which vanish at the critical point, serving as order parameters. The scaling collapses in panels (d) and (f) confirm the scaling relations for these quantities, yielding an estimate of $2\beta/\nu = 1/4$.

Similar to \textit{PCA}, reconstructing input data with a \textit{VAE} enables the detection and characterization of a continuous phase transitions, including those in nonequilibrium systems. For instance, autoencoders have been successfully applied to nonequilibrium models, as discussed in Refs.~\cite{Shen-2021, Shen-2022}. Standard autoencoders have also been used to determine the critical temperature of the Ising model~\cite{Alexandrou-2020}. Moreover, unsupervised learning with \textit{VAEs} provides a powerful approach for analyzing quantum phase transitions~\cite{Luchnikov-2019}.

\section{Conclusions}

In this work, we investigated the MV model and characterized its continuous phase transition using both supervised and unsupervised machine learning techniques. We applied classical methods such as principal component analysis (PCA) alongside deep learning approaches, including classification neural networks and \textit{VAEs}, to estimate the critical points and determine the critical exponents associated with the transition.

Our results demonstrate that machine learning methods can accurately identify transition thresholds and estimate critical exponents, thereby enabling the determination of the universality class of the system. The presented methodologies are general and can be extended to other models of physical interest.

Identifying the universality class is significant, as systems within the same class can be related through non-trivial transformations. Beyond their utility in the study of physical systems, machine learning techniques have broad economic and strategic applications. Thus, their application to academic problems not only advances scientific understanding but also facilitates the broader adoption of these powerful tools.

\section{Acknowledgments} \label{sec:acknowledgements}

We gratefully acknowledge financial support from CAPES (Coordenação de Aperfeiçoamento de Pessoal de Nível Superior), CNPq (Conselho Nacional de Desenvolvimento Científico e Tecnológico), and FAPEPI (Fundação de Amparo à Pesquisa do Estado do Piauí). We thank the \textit{Dietrich Stauffer Computational Physics Lab}, Teresina, Brazil, and the \textit{Laborat\'{o}rio de F\'{\i}sica Te\'{o}rica e Modelagem Computacional (LFTMC)}, Teresina, Brazil, for providing computational resources for the numerical simulations. R. S. Ferreira acknowledges support from FAPEMIG (grant FAPEMIG-APQ-06611-24).

\bibliography{textv1}

\end{document}